# Voltage Asymmetry of Spin-Transfer Torques

Deepanjan Datta [†], Behtash Behin-Aein [†], Sayeef Salahuddin [††], Supriyo Datta [†]

*School of Electrical & Computer Engineering, Purdue University, West Lafayette, IN-47907, USA* [†]

*Department of Electrical Engineering and Computer Science, University of California, Berkeley, CA-94720, USA* [††]

We present a Non-Equilibrium Green's Function (NEGF)-based model for spin torque transfer (STT) devices which provides quantitative agreement with experimentally measured (1) differential resistances, (2) Magnetoresistance (MR), (3) In-plane torque and (4) out-of-plane torque over a range of bias voltages, using a single set of three adjustable parameters. We believe this is the first theoretical model that is able to cover this diverse range of experiments and a key aspect of our model is the inclusion of multiple transverse modes. We also provide a simple explanation for the asymmetric bias dependence of the in-plane torque, based on the polarization of the two contacts in energy range of transport.

Spin Torque Transfer (STT) devices that can switch the magnetization of a soft ferromagnetic layer through spin polarized electrons without any external field have generated significant interest from both basic and applied points of view (See for example Ref. [1-15]). Although the concept of spin torque has been demonstrated by a number of experiments [14,15], quantitative measurement of bias dependence of spin torque has been achieved only very recently [16-20]. Note that these experiments show considerable dispersions in measuring the bias dependence of in-plane torque [16,17,19,20]. Currently, there is no consensus as to a microscopic model that accounts for this discrepancy. Moreover, the existing theoretical models based on effective mass, tight-binding [21-23] and Ab-initio [24,25] band structures do not provide quantitative agreements with the experiments. Therefore, we need a model which can simultaneously explain all of the diverse aspects of STT devices namely i) differential resistances R(V), ii) TMR, iii) in-plane/spin-transfer ($\tau_{\|}$) and iv) out-of-plane/field-like ($\tau_{\perp}$) components of spin torque.

This paper presents a simple effective mass model with five parameters: a) ***equilibrium Fermi level $E_f$***, b) ***Spin-splitting*** $\Delta$, c) ***Barrier height of the insulator $U_b$***, d) ***effective masses for electrons inside FM contacts ($m_{FM,\uparrow}^* = m_{FM,\downarrow}^* = m_{FM}^*$)*** and e) ***effective mass for electrons inside insulator $m_{ox}^*$*** in terms of which we can ***understand all of the aforementioned characteristics (i–iv) of STT devices***. Note that we view $U_b$, $m_{FM}^*$ and $m_{ox}^*$ as parameters that account for a wide variety of factors including imperfection at ferromagnet/insulator interfaces. As such we consider these three parameters adjustable from one structure from another. On the other hand, $E_f$ and $\Delta$ are material parameters. Although this is an effective mass model that does not include bandstructure effects explicitly, we believe that the ***quantitative*** agreement with such a diverse set of experiments shows that it captures much of the essential physics at least in the structures analyzed. One such effect our model tries to capture is the role of transverse modes on the TMR, R(V) and bias asymmetry of the in-plane torque. This last point is currently a topic of debate in the field [16,17,19,20] and our model not only fits the data quantitatively, but also provides a natural explanation for it.

**Theoretical Formalism:** Fig. 1 (a) shows the schematic of a magnetic tunnel junction (MTJ) trilayer device that consists of a fixed magnet with magnetization $\hat{M}$ separated by an insulator from a free layer with magnetization $\hat{m}$, $\theta$ ($\theta = \cos^{-1}(\hat{m}.\hat{M})$) being the angle between them. Current flows along $\hat{y}$, perpendicular to the easy plane ($\hat{x} - \hat{z}$) of both the magnets. Our model (See *supplementary information S1*) for spin transport is based on Non-Equilibrium Green's Function (NEGF) [26-28] method using single band effective mass Hamiltonian described by the parameters described earlier ($\Delta$, $U_b$, $m_{FM}^*$ and $m_{ox}^*$ defined in Fig. 1 (b)). The parameters $E_f$, $\Delta$, and $U_b$ are defined with respect to the bottom of the conduction band of the ferromagnet.

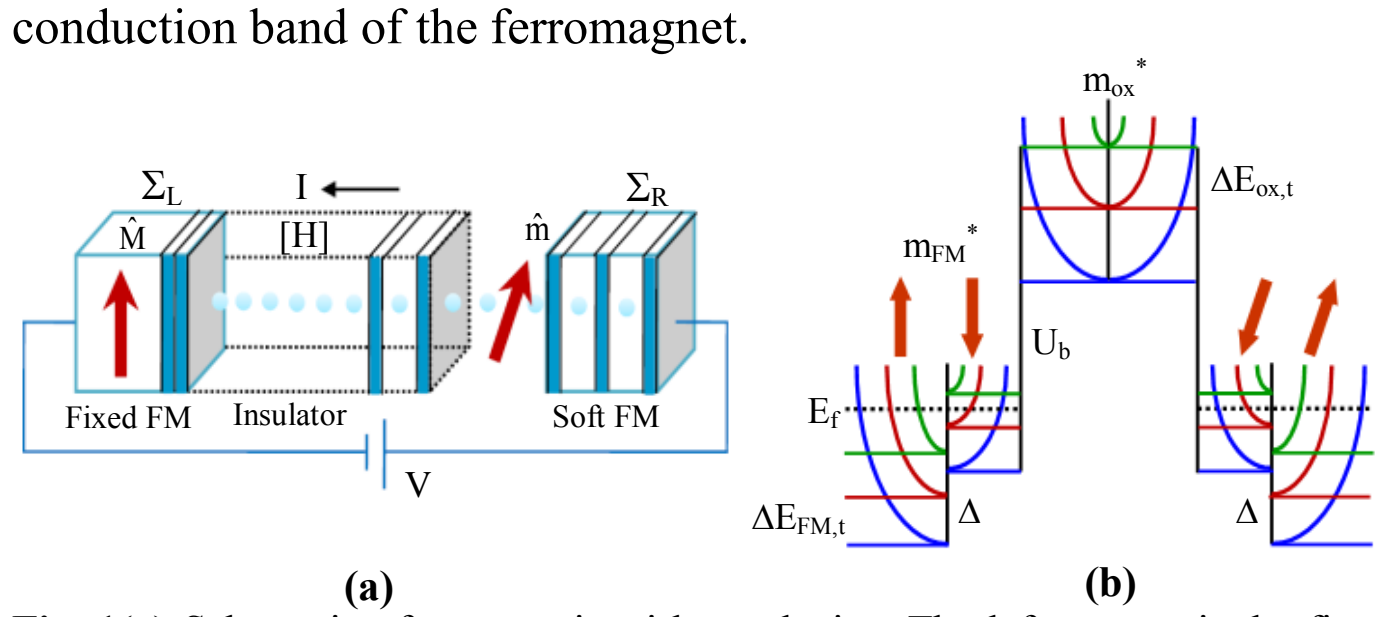

**Fig. 1(a)** Schematic of symmetric tri-layer device. The left contact is the fixed FM having magnetization along the z-axis. The right contact is the free layer and the channel material is an oxide. z is the easy axis and x-z is the easy plane. Transport occurs in y-direction. The device region is modeled using appropriate Hamiltonian H and the contacts by self energy matrices $\Sigma_{L,R}$. **(b)** The figure shows energy-band diagram of the device. The magnets are described by splitted bands of up and down spins while oxide only introduces a barrier. Different transverse modes are shown so that conservation of transverse wave vector is maintained through the device.

The choice of the values of $E_f$ and $\Delta$ are independent of the specific device because in all the experiments CoFeB is used as ferromagnetic contacts. $E_f$ is chosen as 2.25 V, which is believed to be typical for CoFeB alloy [19]. The spin splitting $\Delta$ of 3d states of transition metals is correlated linearly with the spin angular moment $m_S$ (See Fig. 3 of [29]). Based on the experimental studies involving x-ray absorption and magnetic circular dichroism, $m_S/n_{3d}$ = 0.7 (for $Co_{60}Fe_{20}B_{20}$ alloys) [30], where $n_{3d}$ the number of d holes. Since, exact $n_{3d}$ of CoFeB is not known, we use the $n_{3d}$ available for bulk Fe (~3.4) to obtain an $m_S \sim 2.38$ and hence a $\Delta \sim 2.38$ eV. We use a slightly lower value of $\Delta = 2.15$ eV, which is justified due to presence of surface imperfection and other defects at the FM/insulator interface. The effective mass $m_{FM}^*$ of the CoFeB alloy is not reported in literature and is chosen to obtain quantitative agreement for the reported differential resistance, R(V) at different relative angular positions of ferromagnetic contacts. In practice, the values of the effective masses inside ferromagnetic alloys could be very different and might depend on the growth of the crystal inside the lattice. Among the other parameters, we choose barrier height $U_b \sim$ 0.7-1 eV and effective mass of electrons inside insulator $m_{ox}^* \sim$ 0.2-0.3 $m_o$, where $m_o$ is the free electron mass. The range of the values of $U_b$ is motivated by the presence of oxygen vacancy defects that is expected to reduce the effective $U_b$ of MgO tunnel barrier compared to the ideal one (3.7 eV) [31]. The choice of $m_{ox}^*$ also does not differ much from the calculated effective mass of 0.37 $m_0$ from Ab-initio theory [32].

An important aspect of our model is the explicit inclusion of multiple transverse modes with wave vector $k_t$ which are assumed to be decoupled like individual 1-D wires. Device Hamiltonian H and self energy matrices $\Sigma_{L,R}$ for each transverse mode are described by *Eqns. S.I.1* and *S.I.2*. Using H and $\Sigma_{L,R}$, we can calculate the Green's Function G(E) and the spin-dependent correlation function $G^n(E)$, that are used to evaluate charge and spin current densities between different lattice sites for each $k_t$ (See *Eqns. S.I.4, S.I.5; supplementary information S1*). We then sum over the results obtained from all transverse modes to get the final result relevant to realistic experimental devices having sizable cross-sections.

Note that, in our model we assume that the wave vector $k_t$ for each transverse mode is conserved throughout the device. The transverse energies $\Delta E_{FM,t}$ and $\Delta E_{ox,t}$ in the FM region and the oxide (See Fig. 1(b)) are different due to difference in the effective masses. In any practical tunnel-junction structure, the barrier is unlikely to be defect-free over the junction area because of surface imperfections and lattice mismatch between Fe and MgO [32,33]. Such effects could lead to non-conservation of $k_t$, but are outside the scope of this paper.

| Experiments | $E_f$ (eV) | $\Delta$ (eV) | $m_{FM}^*/m_o$ | $U_b$ (eV) | $m_{ox}^*/m_o$ |
|---|---|---|---|---|---|
| **Ref. [16], [17]** | **2.25** | **2.15** | **0.73** | **0.93** | **0.32** |
| **Ref. [19]** | **2.25** | **2.15** | **0.8** | **0.77** | **0.18** |

**Table. 1.** Parameters $E_f$, $\Delta$, $m_{FM}^*$, $U_b$ and $m_{ox}^*$ used to model two sets of available experimental R (V), TMR and Torque.

**R(V) and TMR (Experiment vs. Theory):** Most of the experiments on MTJs measure differential resistance R(V) that is then used to determine TMR ($=(R_{AP}-R_P)/R_P$). Our model successfully describes reported R (V), TMR and its roll-off with voltage, unlike any other theoretical calculation that we are aware of. Fig. 2 (a) and (b) shows comparison between reported [16,17] and calculated R (V) at four different relative angular positions ($\theta$ = $0^o$, $52^o$, $71^o$, and $180^o$) of fixed and free layer magnetizations. The parameters are used to benchmark R (V) data with the experiment are shown in the top row of Table 1.

The simulated bias dependence of R (V) (Fig. 2 (b)) at different angles closely agrees with measured data (Fig. 2 (a)). We see that for the parallel configuration, $R_P$ has very weak dependence on voltage (≈ 3 kΩ). As the angle between the magnetizations of ferromagnetic contacts increases, bias dependence of R(V) becomes stronger. Inset of Fig. 2 (b) shows bias dependence of TMR. Here, the zero voltage TMR (300 K) is calculated to be 150% (indicated by '**O**'), which is very close to the reported value of 154%. We also calculate the expected roll-off of TMR (43% at 0.54V; indicated by '×'). Switching current at parallel configuration is calculated to be 0.64 mA, which is very close to the reported value i.e. 0.6 ± 0.1 mA. Note that the reported switching current is not directly observed, but extrapolated from measured experimental data [16].

We also compared simulated differential resistances with the reported data of Ref. [19] at three different angles $\theta = 0^o$, $137^o$, and $180^o$ (See Fig. 2 (c), (d)). The set of parameters used to benchmark R (V) are shown in the bottom row of Table (1). The values of the above parameters are chosen based on the arguments described above to benchmark the previous experiment. Note that the values for all the parameters except $U_b$ and $m_{ox}^*$ remain unchanged.

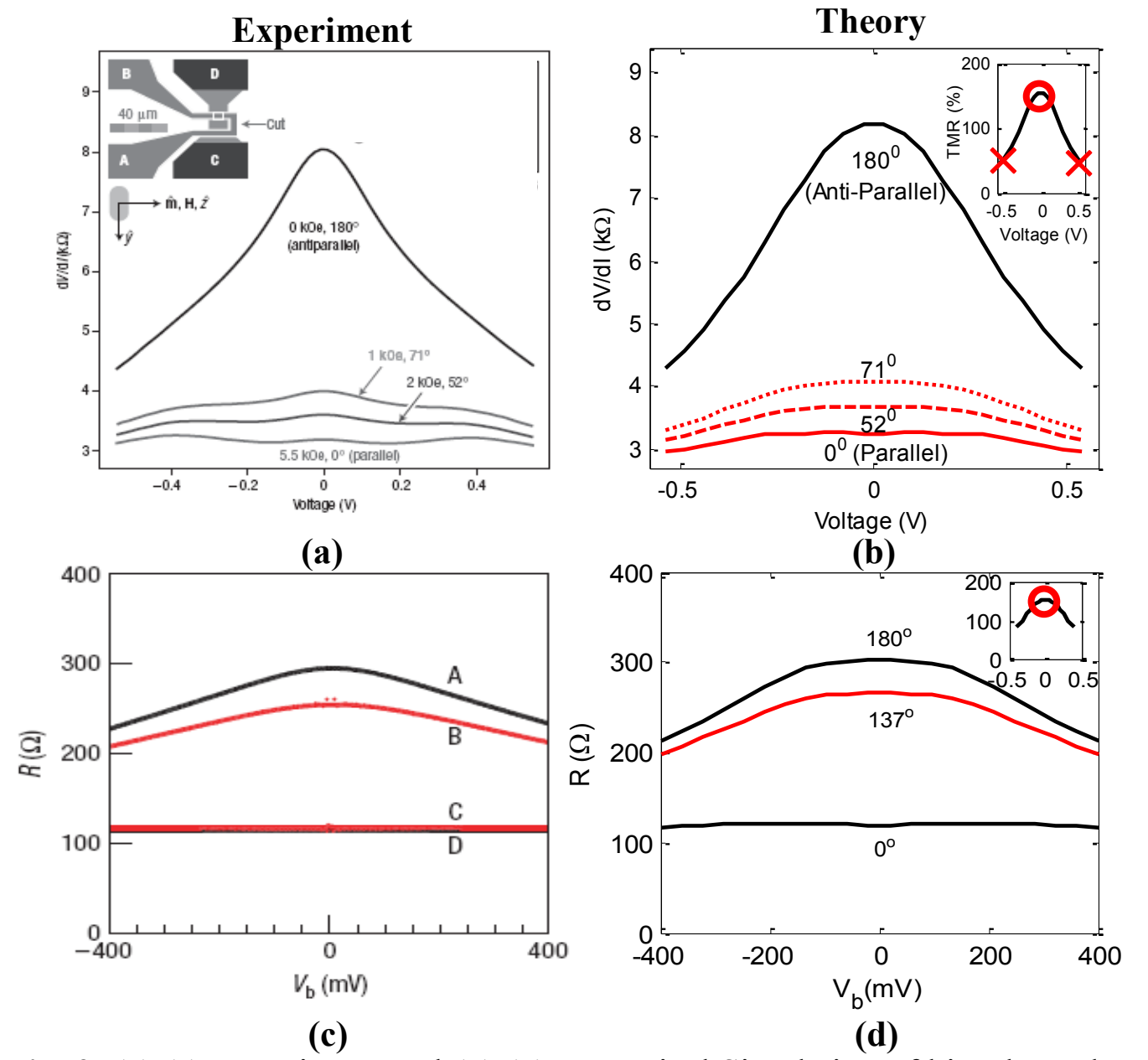

**Fig. 2. (a),(c)** Experiment and **(b),(d)** Numerical Simulation of bias dependence of differential resistances for different angular positions of free FM w.r.t fixed magnet. (Upper panel) [16] and lower panel [19]. Inset of **(b)** and **(d)** shows bias dependence of TMR.

Like the previous experiment, bias dependence of R (V) (See Fig. 2 (d)) closely agrees with those of Fig. 2 (c) for all the angles mentioned above. In parallel configuration, similar to the previous experiment $R_P$ is roughly constant (≈ 122Ω). Inset of Fig. 2 (d) shows bias dependence of TMR. In this case, we calculate the zero voltage TMR (300 K) to be 154% (indicated by '**O**'), which is exactly same as reported in the experiment. Though the experimental switching current density $J_C$ is not reported, based on R(V) curve and the switching voltage for P → AP transition, we can calculate $J_C$ to be $2\times10^7$ A/cm$^2$. The numerical value of $J_C$ is confirmed by micromagnetic simulation as well as Sun's calculation [34] for critical current density.

Note that we obtain close agreement of R (V) and TMR with the measured data from two different experimental groups with a small adjustment of $U_b$ and $m_{ox}^*$.

**Comparison of $\tau_{\|}$ and $\tau_{\perp}$ (or, $\partial\tau_{\|}/\partial V$ and $\partial\tau_{\perp}/\partial V$):** The quantities analyzed above are obtained directly from experiments. However, the reported values of $\tau_{\|}$ and $\tau_{\perp}$ (or, $\partial\tau_{\|}/\partial V$ and $\partial\tau_{\perp}/\partial V$) are not measured directly, rather derived from the measured R.F. voltage output (in the spin-transfer-driven ferromagnetic resonance (ST-FMR) experiments) using a specific theoretical model with its own built-in assumptions. (See Supplementary material [16,17,19]). We calculate torques using the same set of values of the parameters used to benchmark R (V) with the experiment.

Fig. 3(a) and (b) show comparison between experimental [15,16] and calculated $\partial\tau_{\|}/\partial V$ and $\partial\tau_{\perp}/\partial V$ respectively at two different relative angular positions ($\theta$=58$^o$ and 131$^o$) of fixed and free layer magnetizations. From our numerical simulation, we see that both the torkances are proportional to Sinθ, so when these quantities are normalized with respect to Sinθ, they should fall onto single curves (See Fig. 3 (b)). It confirms the angular dependence of both the torkances. Note that the reported data [17] shows 16% variation of in-plane torkance for |V|≤0.3V) and concluded the earlier measured flat nature of $\partial\tau_{\|}/\partial V$ [16] as an artifact of neglecting significant terms in the derivation of the RF voltage output in ST-FMR experiment. Upper panel of Fig. 3(b) shows the calculated bias dependence of in-plane torkance for the bias range of $|V|\leq 0.3V$, which agrees both qualitatively and quantitatively with measured data. With regards to the absolute magnitudes of $\partial\tau_{||}/\partial V$, our simulation shows that near V=0, $(\partial\tau_{||}/\partial V)/\mathrm{Sin}\theta$ is 0.11(ħ/2q) $k\Omega^{-1}$, which is in agreement with the reported data i.e. (0.1(ħ/2q) $k\Omega^{-1}$) [17] and also with the Ab-inito study (0.14(ħ/2q) $k\Omega^{-1}$) [23]. In contrast to the in-plane component of torkance, the bias dependence of the out-of-plane component $\partial\tau_{\perp}/\partial V$, shows $\partial\tau_{\perp}/\partial V \propto V$ (normalized with respect to Sinθ), so that $\tau_{\perp} = (A_0+A_1V^2)\ \mathrm{Sin}\theta$. Also from the lower panel of Fig. 3(b), it is evident that $A_1$ is completely independent of θ for the tunneling devices.

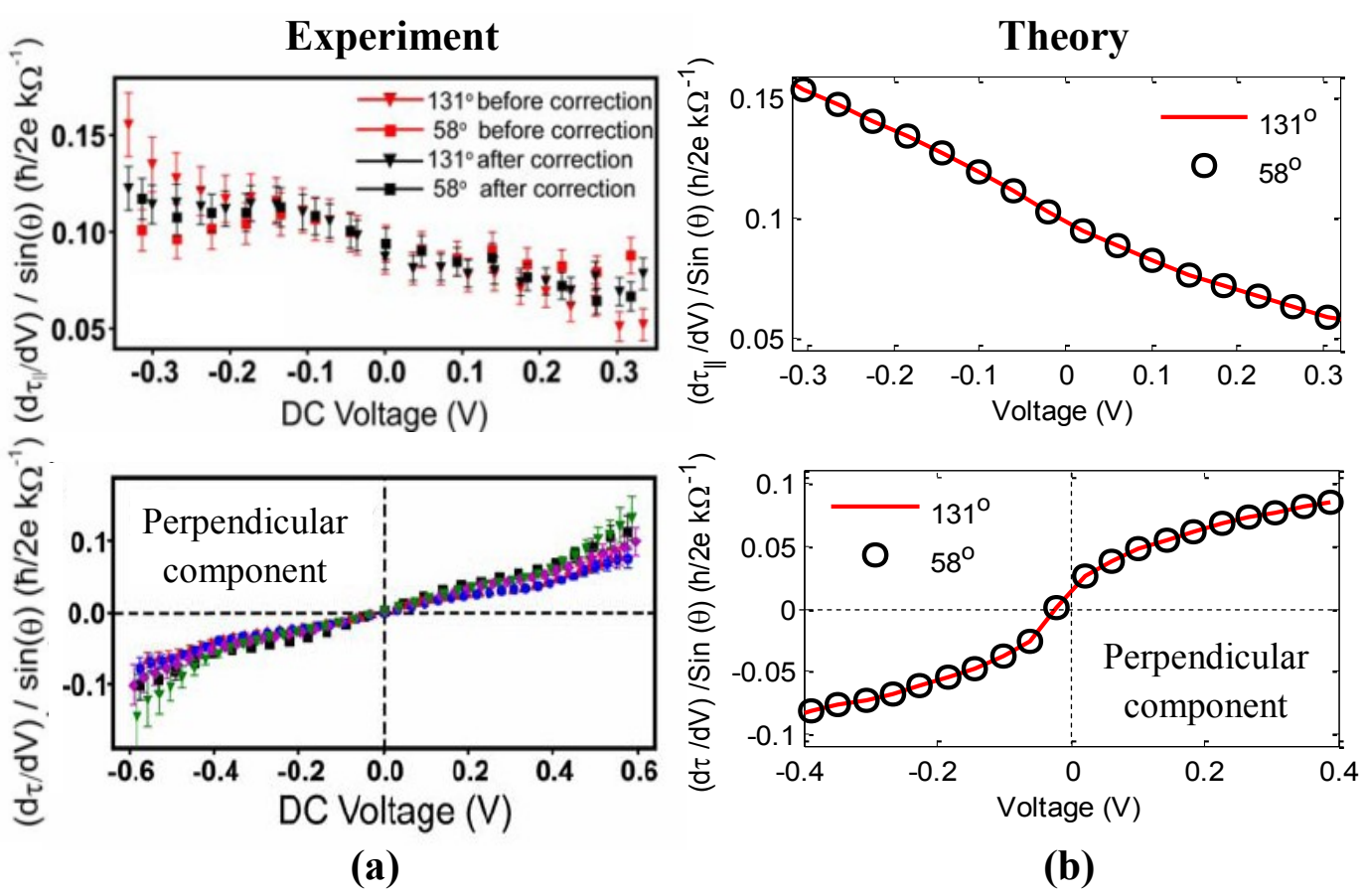

**Fig. 3. (a)** Measured (Cornell) [17] and **(b)** simulated bias dependence of in-plane and out-of-plane torkances are plotted for two different angles between the fixed and soft magnets i.e. 58° and 131°.

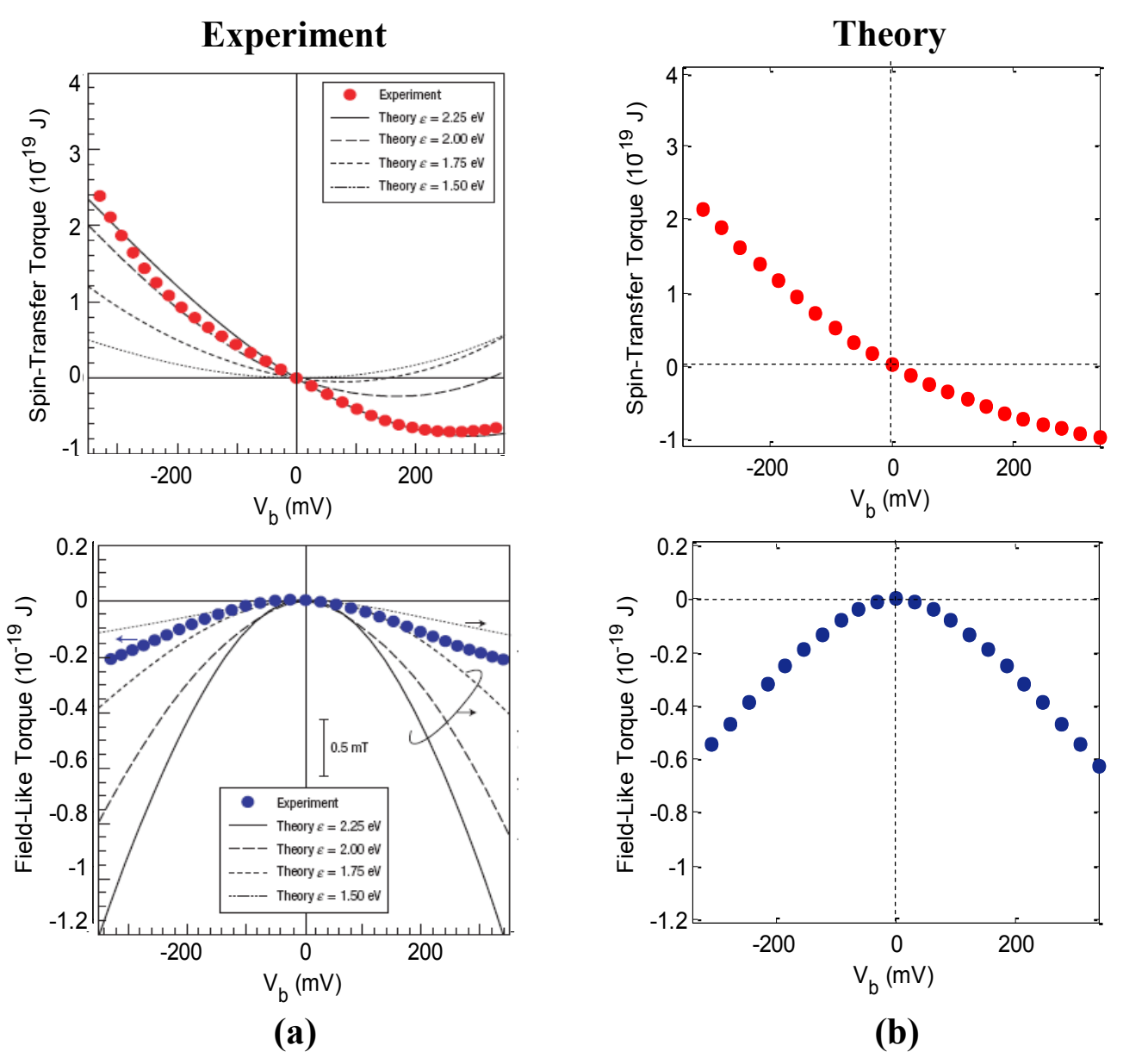

**Fig. 4. (a)** Measured (Tsukuba) [19] and **(b)** simulated bias dependence of magnitudes of the simulated in-plane (Spin-Transfer) and out-of-plane (Field-like) torques.

The magnitudes of $\tau_{||}$ and $\tau_{\perp}$ reported by Ref. [19] is compared with the calculated torques as shown in Fig. 4 (a), (b). We use same set of values of the parameters used to benchmark calculated R (V) data with the experiment. From our simulation, we see that spin transfer component of the torque both quantitatively and qualitatively matches with the reported value. There is some discrepancy in the magnitude of the field like torque $\tau_{\perp}$, but remember that the reported data is based on an indirect measurement of torque as reported in literature. Likewise the experiment, we plot only the non-equilibrium portion of $\tau_{\perp}$ (i.e. $A_1V^2$). It is important to note that the magnitude of the equilibrium part i.e. $A_0$ is comparable to $A_1V^2$ at lower voltages and can manifests itself as an exchange field introducing an asymmetry in the R-H loops.

Now, one can obtain the simulated switching torque amplitudes from Fig. 4(a) and (b). Although the switching voltages for both AP → P (-270 mV) and P → AP (380 mV) transitions have been reported, the reported torque values are available only for AP → P switching. We have performed micromagnetic simulations to confirm that the torque magnitude from the experiment and our model do result in switching and are in agreement with the threshold values of switching torque [35].The details of micromagnetic simulation are discussed in *Supplementary information S2*.

**Conditions for Asymmetry:** As remarked in the introduction, the bias asymmetry of $\tau_{||}$ (as well as $\partial\tau_{||}/\partial V$) has caused much discussion in the literature, since many theoretical models [2,24,25] predict that $\tau_{||}(V)$ should be perfectly anti-symmetric (~aV), making $\partial\tau_{||}/\partial V$ perfectly anti-symmetric with V, but the results in Fig. (3), (4) indicate otherwise. To gain insight, we use our own numerical models to look at the quantity $(\tau_{||}(V)+\tau_{||}(-V))/\ (|\tau_{||}(V)|+|\tau_{||}(-V)|)$ for different transverse modes individually as shown in Fig. 5(b), where the x-axis is the voltage V and the y-axis is $\Delta$ - $(E_f-E_t)$. Note that $(\tau_{||}(V)+\tau_{||}(-V))/\ (|\tau_{||}(V)|+|\tau_{||}(-V)|)$ is essentially zero as long as

$$\Delta\text{-}\left(E_f\text{-}E_t\right) > q|V| \quad (1)$$

We have found the above criterion to hold for all choices of $E_f$ and $\Delta$ so that $\Delta$-$E_f$ remains constant. One can see from Fig. 5(a) that the transverse modes, for which this inequality is satisfied, effectively see "*half-metallic*" contacts on either side. This is because for any transverse mode $E_t$, $(E_f-E_t)$ represents the maximum longitudinal energy, so that modes satisfying Eqn. (1) (which we call type '**A**') only one type of spin is injected and detected and it seems that this condition leads to perfect anti-symmetry. In an actual device we have contributions from both type '**A**' modes as well as other modes where either one type of spin is injected but both spins are detected (type '**B**') or both spins are injected and/or, detected (type '**B′**' modes). Neither type '**B**' nor '**B′**' mode satisfies Eqn. (1), and the degree of asymmetry actually observed in the experimental devices suggests that their contribution to $\tau_{||}$ dominate that from type 'A' mode.

**Reason for Asymmetry:** Interestingly, both the experiments and our numerical results show that magnitude of the torque $\tau_{||}(V)$ is smaller when a negative voltage is applied to the free layer (or, positive voltage to fixed layer as shown upper panel of Fig. 4) so that free layer bands are raised up. Asymmetry in the magnitude of $\tau_{||}$ is explained by Fig. 5(a). Let us assume the fixed layer is along $\hat{x}$-direction, while free layer is along $\hat{z}$-direction. Therefore, according to Eqn. (S.I.6) (*supplementary information S1*), in-plane torque on the free layer is $\tau_{||,\ \mathrm{free}}$ (per Bohr Magneton) = $I_{Sz}$; while torque acting on the fixed layer is $\tau_{||,\ \mathrm{fixed}}$ (per Bohr Magneton) = -$I_{Sx}$. Now, applying a negative voltage to free layer is the same as applying positive voltage to fixed layer and we expect torque exerted on the free layer at negative voltage to be the same as that exerted on the fixed layer at positive voltage. This is indeed true as evident from the $I_{Sz}$ and $I_{Sx}$ plots shown in Fig. 5(c).

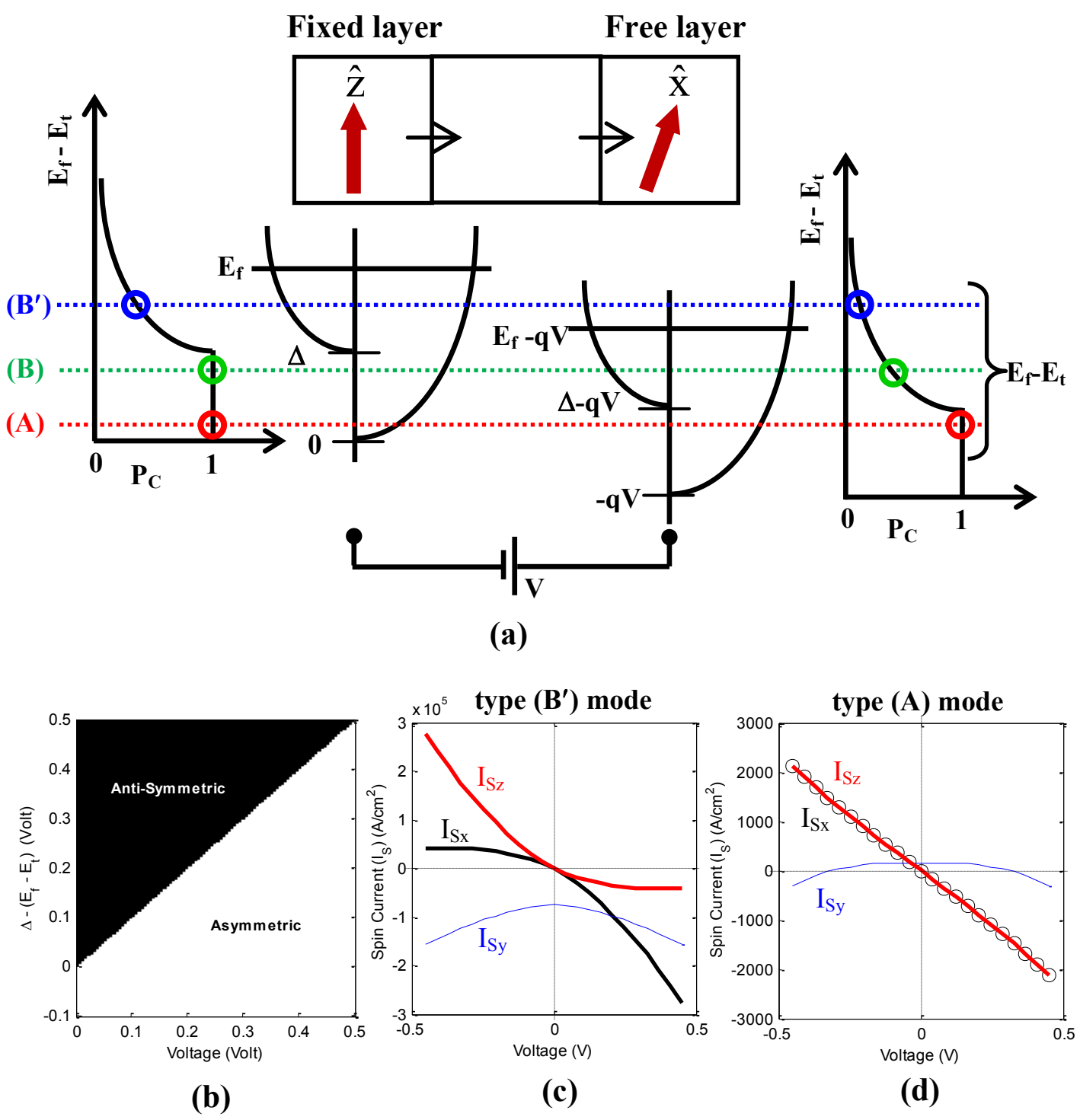

**Fig. 5. (a)** The schematic shows the fixed and free layers are along z and x-axis respectively. Ferromagnet band-structure is drawn explaining the relative contribution of transverse modes on the asymmetric bias dependence of $\tau_{||}$. ($E_f$-$E_t$) represents maximum longitudinal energy for which there are three types of modes: type (**A**) ($E_f$-$E_t$<Δ-qV), type (**B**) (Δ-qV< $E_f$-$E_t$<Δ), and (**B′**) modes ($E_f$-$E_t$>Δ). $E_f$-$E_t$ vs. $P_C$ is drawn for both the FM contacts showing relative contributions of density of states of majority and minority carriers available for tunneling. Parameters used to generate **(b)**, **(c)** and **(d)** are same as shown in the top row of Table 1. **(b)** shows the plot of the function $(\tau_{||}(V)+\tau_{||}(-V))/(|\tau_{||}(V)|+|\tau_{||}(-V)|)$ for different transverse modes, the x and y axes are being voltage V, and Δ-($E_f$-$E_t$). Black denotes a value of the function less than a part per million ($10^{-6}$). It separates the regions of anti-symmetric and asymmetric bias dependence of $\tau_{||}$. The two variables V and Δ- ($E_f$ -$E_t$) have been varied independently to produce 2D grid with 2,50,000 grid points. **(c)** Spin current is calculated for type (**B′**) modes; When negative voltage is applied to fixed layer, we get $I_{Sz}$>$I_{Sx}$. When the voltage polarity is reversed, we get $I_{Sz}$<$I_{Sx}$. Bias dependence of spin currents for type (**B**) modes is same as that of (**B′**) modes. **(d)** Spin current is calculated for type (**A**) modes. We see that $I_{Sz}$ = $I_{Sx}$ irrespective of voltage polarity. We see that the magnitude of spin current is much greater for both type (**B**) and (**B′**) modes compared to type (**A**) modes. Therefore, type (**B**) and (**B′**) modes contribute more in tunneling compared to type (**A**) modes for the devices explored in the literatures.

Note that $I_{Sz}$ > $I_{Sx}$ (for type '**B′**' (as well as '**B**') modes), when negative voltage is applied to fixed ($\hat{z}$) magnet w.r.t. free ($\hat{x}$) magnet (Fig. 5(c)). But $I_{Sz}$ = $I_{Sx}$ for type '**A**' modes. This can be understood by considering spin polarization ($P_C$) of the fixed ($\hat{z}$) magnet and free ($\hat{x}$) magnets in the energy range where current flow occurs for a given transverse mode (See Fig. 5(a)).

| Type of Modes | Energy range of transport | $P_C$ ($\hat{z}$) | $P_C$ ($\hat{x}$) | $\tau_{\|\|}$ (V) |
|---|---|---|---|---|
| **A** | $E_f$-$E_t$ < Δ-qV | 1 | 1 | Anti-symmetric |
| **B** | Δ-qV <$E_f$-$E_t$ <Δ | 1 | < 1 | Asymmetric |
| **B′** | $E_f$-$E_t$ > Δ-qV | < 1 | < 1 | Asymmetric |

**Table. 2.** The modes corresponding to particular energy range of transport and bias dependence of in-plane torque,$\tau_{||}$.

When the polarity of the voltage is reversed, for the type '**B′**' modes, $P_C$ ($\hat{z}$) <$P_C$ ($\hat{x}$) <1 and for the '**B**' modes, $P_C$ ($\hat{z}$) <$P_C$ ($\hat{x}$) = 1. In both cases, $I_{Sz}$ < $I_{Sx}$ (See Fig. 5(c)). For type '**A**' modes, $P_C$ ($\hat{z}$) = $P_C$ ($\hat{x}$) =1 and hence, $I_{Sz}$=$I_{Sx}$ (Fig. 5(d)). Note that the out-of-plane component is determined by $I_{Sy}$ whose magnitude depends on the $P_C$ of both the contacts and hence unaffected by the voltage polarity. By contrast the in-plane torque on the free layer is determined by $I_{Sz}$ whose magnitude depends on the $P_C$ of the fixed ($\hat{z}$) magnet and is larger when a negative voltage is applied to it. This seems to provide a simple explanation for the asymmetry of the in-plane torque (top panel of Fig. 3, 4) which attracted much discussion.

To summarize, we have described a quantum transport model for a spin torque device using NEGF formalism, which provides both qualitative and quantitative agreement on differential resistances R(V), TMR, and torkances simultaneously with a single set of parameters (i.e. effective masses, barrier height and spin-splitting in contacts). We show that bias asymmetry in $\tau_{||}$ can be understood in terms of the polarization ($P_C$) of the fixed layer in the energy range of transport. Though our analysis is focused on vertical STT devices, the general formalism can also be used for analyzing bias dependence of spin torque in non-local lateral spin-torque devices [36,37].

The work is supported by MARCO focus center for Materials, Structure, and Devices.

# Supplementary Information

## Voltage Asymmetry of Spin-Transfer Torques

Deepanjan Datta[†], Behtash Behin-Aein[†], Sayeef Salahuddin[††], and Supriyo Datta[†]

*School of Electrical & Computer Engineering, Purdue University, West Lafayette, IN-47907, USA* [†]

*Department of Electrical Engineering and Computer Science, University of California, Berkeley, CA-94720, USA* [††]

### S1. Details of Non-Equilibrium Green's Function (NEGF) Model used

The Non Equilibrium Green's Function (NEGF) based transport model for MTJ devices is based on single band effective mass Hamiltonian H and self-energy $\Sigma_{L,R}$ which are used to calculate Green's function G(E), electron correlation function $G^n$, charge and spin current densities J and $J^S$ respectively.

#### (a) Hamiltonian and Self-Energy matrix

Here, we will present the device Hamiltonian H and self energy matrices $\Sigma_{L,R}$ for each transverse mode $k_t$ through the device. We assume that the fixed magnet $\hat{M}$ and the soft magnet $\hat{m}$ are both in the $\hat{x}$ - $\hat{z}$ plane (see Figure S1). Hamiltonian for each transverse mode is given by ( $\hat{\sigma} \equiv$ Pauli Spin matrices)

***Left Contact:***

$$H_L(i,j,k_t) = \begin{cases} \left(\alpha_{FM}(k_t) + qV/2\right) I + \left(\dfrac{I - \hat{\sigma}.\hat{M}}{2}\right)\Delta, & i = j \\ -t_{FM} I, & \text{if } i \text{ and } j \text{ are nearest neighbors} \\ 0, & \text{otherwise} \end{cases} \quad \text{(S.I.1. a)}$$

***Left Interface:***

$$H_{interface}(i,i,k_t) = \left(\alpha_{int}(k_t) + \left(U_b/2\right) + \left(qV/2\right)\right)I + \left(\frac{I - \hat{\sigma}.\hat{M}}{4}\right)\Delta \quad \text{(S.I.1. b)}$$

***Channel:***

$$H_{channel}(i,j,k_t) = \begin{cases} \left(\alpha_{ox}(k_t) + U_b + qV\left(\dfrac{1}{2} - \dfrac{i}{N+1}\right)\right) I, & i = j \\ -t_{ox} I, & \text{if } i \text{ and } j \text{ are nearest neighbors} \\ 0, & \text{otherwise} \end{cases} \quad \text{(S.I.1. c)}$$

***Right Interface:***

$$H_{interface}(i,i,k_t) = \left(\alpha_{int}(k_t) + \left(U_b/2\right) - \left(qV/2\right)\right)I + \left(\frac{I - \hat{\sigma}.\hat{m}}{4}\right)\Delta \quad \text{(S.I.1. d)}$$

$$\textbf{\textit{Right Contact:}} \quad H_R(i,j,k_t) = \begin{cases} \left(\alpha_{FM}(k_t) - qV/2\right) I + \left(\dfrac{I - \hat{\sigma}.\hat{m}}{2}\right)\Delta, \; i=j \\ -t_{FM}\, I, \text{ if } i \text{ and } j \text{ are nearest neighbors} \\ 0, \text{ otherwise} \end{cases} \tag{S.I.1. e}$$

where, $\alpha_{FM}(k_t) = 2t_{FM} + \hbar^2 k_t^2 / 2m_{FM}^*$, $\alpha_{ox}(k_t) = 2t_{ox} + \hbar^2 k_t^2 / 2m_{ox}^*$, $\alpha_{int}(k_t) = 0.5(\alpha_{FM} + \alpha_{ox})$, $t_{FM} = \hbar^2 / 2m_{FM}^* a^2$, $t_{ox} = \hbar^2 / 2m_{ox}^* a^2$, $m_{FM}^*$ and $m_{ox}^*$ are effective masses of electron inside FM and insulator region respectively and $a$ is the uniform lattice spacing and $I$ is the $2\times2$ identity matrix. $N$ is the total number of atomic sites (excluding the interface points) inside the insulator. $V$ is the applied voltage difference between two FM contacts which is assumed to drop linearly inside the insulator. $U_b$ is the barrier height of the insulator.

To write the self-energy matrices $\Sigma_{L,R}$ we note that their non-zero elements are the left (L) or, right (R) lattice points respectively and these are given by a $2\times2$ matrix of the form

$$\Sigma_{L,R}(i,i,k_{\Box}) = \begin{bmatrix} -t_{FM}\exp\left(i\,k_{L,R}^{\uparrow}a\right) & 0 \\ 0 & -t_{FM}\exp\left(i\,k_{L,R}^{\downarrow}a\right) \end{bmatrix} \tag{S.I.2. a}$$

where,

$$k_{L,R}^{\uparrow}a = \cos^{-1}\left(1 - \frac{E \mp qV/2 - \left(h^2 k_t^2 / 2m_{FM}^*\right)}{2t_{FM}}\right) \tag{S.I.2. b}$$

$$k_{L,R}^{\downarrow}a = \cos^{-1}\left(1 - \frac{E \mp qV/2 - \left(h^2 k_t^2 / 2m_{FM}^*\right) - \Delta}{2t_{FM}}\right) \tag{S.I.2. c}$$

(Use $+qV/2$ for the contact L, and $-qV/2$ for the contact R) and $E$ is energy. The form given above for $\Sigma_{L,R}(i,i,k_t)$ is appropriate, if the easy axis of the magnet is used as the quantization axis for the spin. If the easy axis does not lie along $\hat{z}$, then a unitary transformation is needed to transform this quantization axis to $\hat{z}$ [S1].

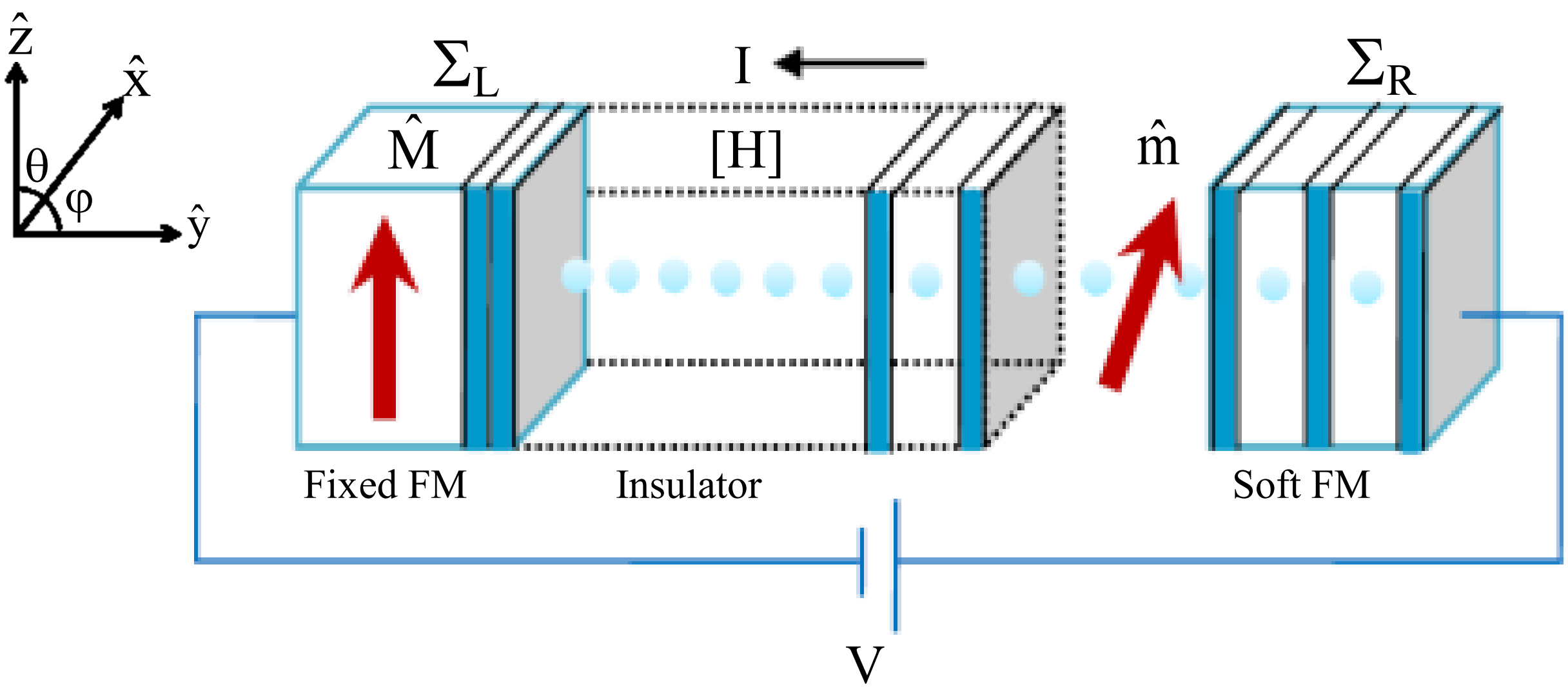

**(a)**

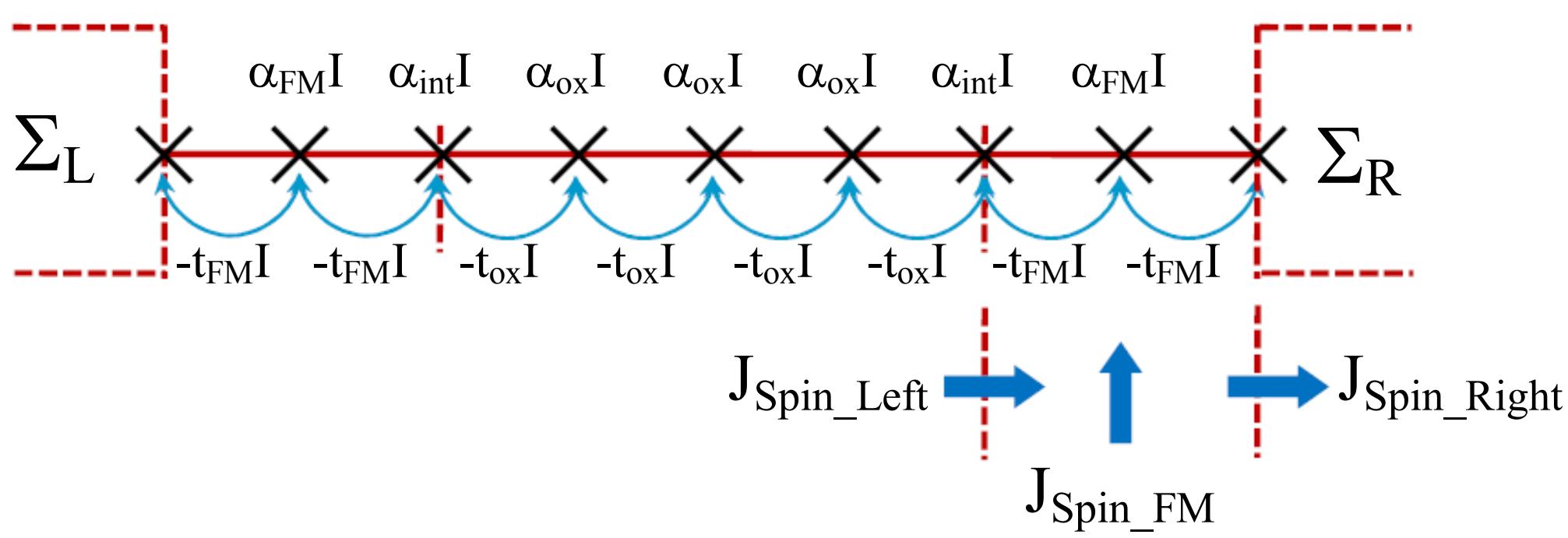

**(b)**

**Figure S1**. **(a)** The schematic of FM/Insulator/FM device is shown. The device region is modeled using appropriate Hamiltonian [H] and the contacts are taken into account by self energy matrices $\mathbf{\Sigma_{L,R}}$ whose anti-Hermitian components $\Gamma_{L,R} = i(\mathbf{\Sigma_{L,R}} - \mathbf{\Sigma_{L,R}}^{\dagger})$ are broadening matrices due to contacts L and R respectively. [2,3] **(b)** $t_{FM}$ and $t_{ox}$ represent the coupling parameter between each lattice site given by $t_{FM} = \hbar^2/2m_{FM}^*a^2$, $t_{ox} = \hbar^2/2m_{ox}^*a^2$, a being uniform lattice spacing in discrete representation. $\alpha_{FM}$ and $\alpha_{ox}$ are given by $\alpha_{FM} = 2\ t_{FM} + \hbar^2k_t^2/2m_{FM}^*$ and $\alpha_{ox} = 2\ t_{ox} + \hbar^2k_t^2/2m_{ox}^*$, where, $\hbar^2k_t^2/2m_{FM}^*$ and $\hbar^2k_t^2/2m_{ox}^*$ represent transverse energy of a mode with wave vector $k_t$ inside FM and insulator regions respectively. $J_{Spin_Left}$ and $J_{Spin_Right}$ are current densities going into and coming out of the soft ferromagnetic region. Therefore, the current density absorbed is $J_{Spin_FM} = J_{Spin_Left} - J_{Spin_Right}$, which exerts torque on the soft FM layer.

**(b) Green's Function, Charge and Spin Current Density**

In Appendix **S1(a)**, we showed the device Hamiltonian H and self-energy matrices $\Sigma_{L,R}$ for the device. Once H and $\Sigma_{L,R}$ are known, all quantities of interest including Green's function G(E) can be calculated from the following set of equations [S2, S3]

$$\begin{aligned}
&\textit{\textbf{Green's Function:}} && G(E) = \left[E - H - \Sigma_L - \Sigma_R\right]^{-1};\\
&\textit{\textbf{Spectral Density:}} && A = i\left(G - G^\dagger\right) = G\,\Gamma\,G^\dagger;\\
&\textit{\textbf{Electron Correlation Function:}} && G^n(E) = G\left(\Sigma_L^{in} + \Sigma_R^{in}\right)G^\dagger;\\
&\textit{\textbf{In-scattering Function:}} && \Sigma_{L,R}^{in}(E) = \Gamma_{L,R}(E)\ f_{L,R}(E);\\
&\textit{\textbf{Broadening Matrix:}} && \Gamma_{L,R}(E) = i\left(\Sigma_{L,R}(E) - \Sigma_{L,R}^\dagger(E)\right);
\end{aligned} \tag{S.I.3}$$

where, $G^n$ ($\equiv -iG^<$) refers to spin-dependent correlation function whose diagonal elements are electron density. A is the spectral function, whose diagonal elements are local density of states, $\Sigma^{in}$ is the in-scattering function representing the rate at which electrons come in to the device from the contacts, and $f_{L,R}$ are the Fermi functions in contacts 1 and 2 respectively. Once we know $G^n$, the charge and spin current density between lattice sites are calculated as [S3]

***Charge Current Density:***
$$J_{j,j\pm1} = \frac{1}{i\hbar}\int_E dE\ \mathrm{Real}\left[\mathrm{Trace}\left(H_{j,j\pm1}G^n_{j\pm1,j} - G^{n\dagger}_{j,j\pm1}H^\dagger_{j\pm1,j}\right)\right] \tag{S.I.4}$$

***Spin Current Density:***
$$J^S_{j,j\pm1} = \frac{1}{i\hbar}\int_E dE\ \mathrm{Real}\left[\mathrm{Trace}\left\{\hat{\sigma}.\left(H_{j,j\pm1}G^n_{j\pm1,j} - G^{n\dagger}_{j,j\pm1}H^\dagger_{j\pm1,j}\right)\right\}\right] \tag{S.I.5}$$

The spin torque exerted on the free magnetic layer will be the difference between spin current going into and coming out of the soft magnet in the absence of scattering. We define spin torque as the divergence of spin current [S4]

$$\vec{\tau} = -\mu_B\int_\Omega dV\ \left[\vec{\nabla}.\vec{J}_S\right] = \mu_B\int_S dS\ \left(-\int dy\ \vec{\nabla}.\vec{J}_S\right) = \mu_B\int_S dS\ \left(\vec{J}_{Spin_Left} - \vec{J}_{Spin_Right}\right) = \mu_B\int_S dS\ \vec{J}_{Spin_FM} \tag{S.I.6}$$

where, S and *V* are the area of cross-section and the volume of magnet respectively. $\mu_B$ is the Bohr Magneton (i.e. magnetic moment per unit volume) and $J_S$ is the spin current density between two lattice points. In our calculations, however, $\int_S dS\ \vec{J}_{Spin_FM}$ is replaced by $\int_S dS\ \vec{J}_{Spin_Left}$ based on following two assumptions: (i) Spin-polarized electrons becomes completely polarized along the free layer magnetization $\hat{m}$ after traversing a few monolayer inside the magnet. (ii) Spin torque acting on the free layer has components only transverse to the free layer magnetization.

## S2. Micromagnetic Simulation

Current induced magnetization switching in Spin Torque Transfer (STT) devices is of great interests because of its range of potential applications. To integrate the STT devices with existing semiconductor technology, lower switching current densities are very important. The current density for switching the magnetization of the free layer is derived by Sun [S5]

$$J_C = \frac{\alpha}{\eta}\left(\frac{2e}{\hbar}\right)\left(t_{FM} H_C M_S\right)\left(1+\frac{2\pi M_S}{H_C}+\frac{H}{H_C}\right) \quad \text{(S.II.1)}$$

The corresponding condition for threshold value for the switching torque is [S5]

$$\frac{H_{S,min}}{H_C} = \alpha\left(1+\frac{2\pi M_S}{H_C}+\frac{H}{H_C}\right) \quad \text{(S.II.2)}$$

where, $H_{S,min}$ is the minimum field required for switching (due to spin torque), $H_C$ is the uniaxial anisotropy field, and $\alpha$ is the damping parameter. The effect of demagnetizing field ($H_d$) is included by the term $2\pi M_S/H_C$. H includes the external magnetic field and the field due to out-of-plane component of spin torque. $t_{FM}$ is the thickness of the thickness of the free FM layer. Now, $H_C$ and demagnetizing field of the free layer are 45 Oe and 1.38 T respectively [S6]. Therefore, with applying any external field and assuming the damping coefficient $\alpha$ to be 0.008, the condition for switching becomes $H_{S,min}$ = 55.8 – 0.008 $H_{out\text{-}of\text{-}plane}$.

Now, we will describe the magnetization dynamics with macrospin approximation. If spin torque $\vec{\tau}$ is exerted on the free layer with magnetization $\hat{m}$, the rate of change of $\hat{m}$ is described by landau-Lifshitz-Gilbert's Eqn. and is written as,

$$\frac{d\hat{m}}{dt} = -|\gamma|\ \hat{m}\times\vec{H}_{eff} + \alpha\ \hat{m}\times\frac{d\hat{m}}{dt} + \vec{\tau} \quad \text{(S.II.3)}$$

where, $\vec{H}_{eff}\left(=\vec{H}_{ext}+\vec{H}_C+\vec{H}_d\right)$ can be written as the gradient of the potential energy w.r.t. normalized magnetization i.e. $\vec{H}_{eff} = -\left(1/M_S V\right)\ \vec{\nabla}_m E$ where $M_S$ is the saturation magnetization. The above equation can be simplified as,

$$\left(1+\alpha^2\right)\frac{d\hat{m}}{dt} = -|\gamma|\ \hat{m}\times\left(\vec{H}_{eff} - \frac{\alpha}{|\gamma|}\ \vec{\tau}\right) - \alpha\ \hat{m}\times\left(\hat{m}\times\vec{H}_{eff}\right) + \vec{\tau} \quad \text{(S.II.4)}$$

We include the effect of spin torque by defining the components of $\vec{\tau}$ as $\vec{\tau}_\perp = \tau_\perp(V)\left(\hat{m}\times\hat{M}\right)$ and $\vec{\tau}_\parallel = \tau_\parallel(V)\ \hat{m}\times\left(\hat{m}\times\hat{M}\right)$ . Substituting in Eqn. (S.II.4), we get

$$\left(1+\alpha^2\right)\frac{d\hat{m}}{dt} = -|\gamma|\ \hat{m}\times\left(\vec{H}_{eff}+A\,\hat{M}\right) - \alpha|\gamma|\ \hat{m}\times\left(\hat{m}\times\left(\vec{H}_{eff}+B\,\hat{M}\right)\right) \quad \text{(S.II.5)}$$

where, the following notations have been adopted:

$$A \equiv \tau_{\perp}(V) - \alpha\,\tau_{\parallel}(V) \qquad\qquad B \equiv \tau_{\perp}(V) + \tau_{\parallel}(V)/\alpha$$

Now, we solved the Eqn. (S.II.5) with varying the magnitude of $\tau_{||}(V)$ and $\tau_{\perp}$ (V) calculated for different voltages. We see that AP → P switching occurs when a negative voltage is applied to the fixed layer (V = -0.27 V) which is exactly same as the switching voltage mentioned in Ref. [S6]. We see that at V = -0.27 V, magnitude of the in-plane torque is $H_{||}(V=-0.27V) = \tau_{||}(V=-0.27V)/M_SV = 60$ Oe and $H_{\perp}(V=-0.27V) = \tau_{\perp}(V=-0.27V)/M_SV = 40$ Oe. Now if we put the values of $H_{\perp}$ in Eqn. (S.III.2), $H_{S,min} = 55.8$ Oe $< H_{||}(V=-0.27V)$. Therefore, the condition for switching derived by Sun is satisfied with the values we obtained from NEGF simulation. Figure S2 shows the magnetization vs. time diagram of our micromagnetic simulation.

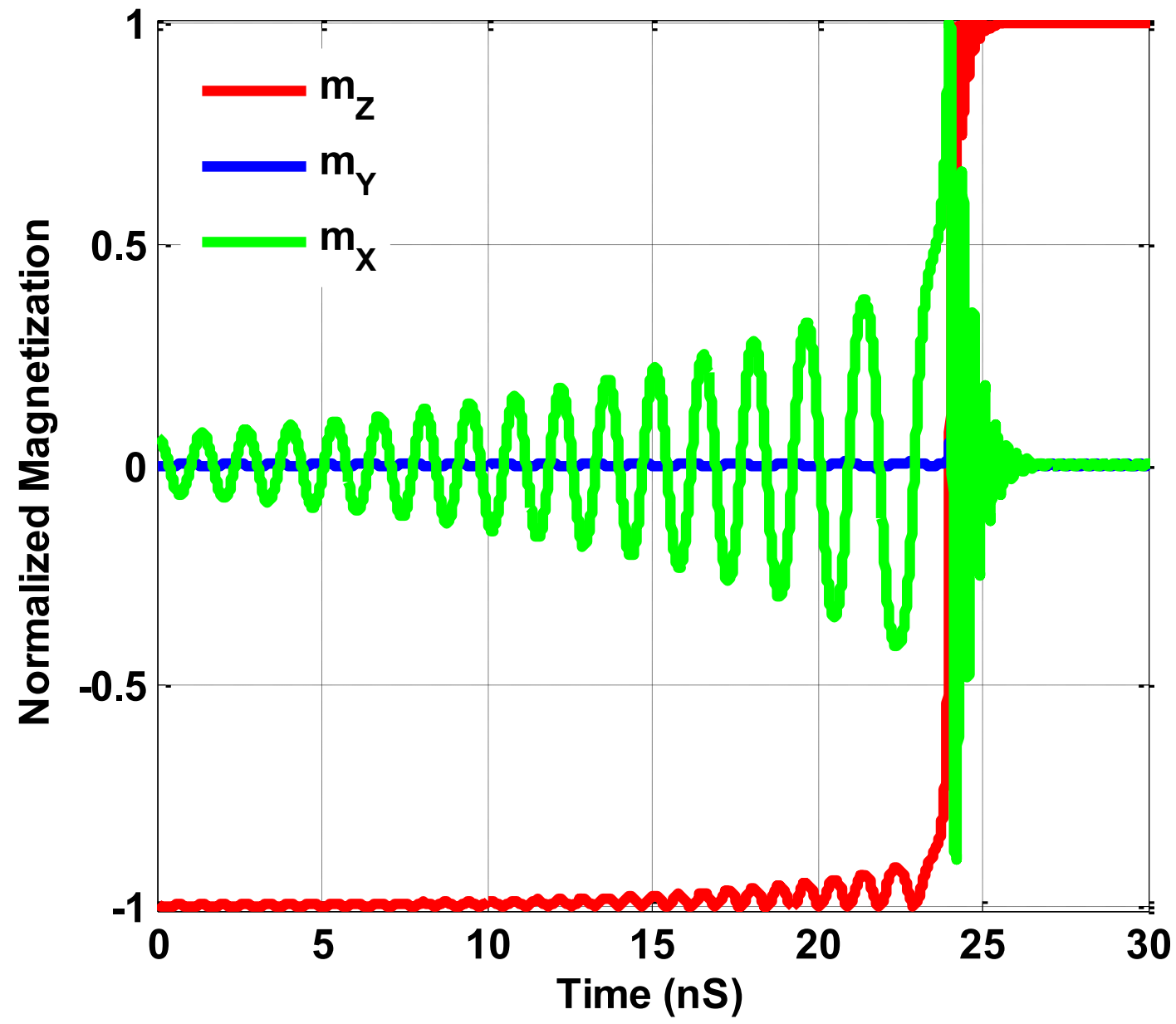

**Figure S2**. Micromagnetic simulation for AP → P switching. The parameters used are: $H_C$ = 45 Oe, $H_{\|}(V=-0.27V) = \tau_{\|}(V=-0.27V)/M_SV$ = 60 Oe and $H_{\perp}(V=-0.27V) = \tau_{\perp}(V=-0.27V)/M_SV$ = 40 Oe, $M_S$ = 1100 emu/ $cm^3$, $t_{FM}$ = 2 nm and $\alpha$ = 0.008. The switching occurs exactly at V = -0.27 V which is reported in paper. [S5] For this simulation we have taken into account the out-of-plane demagnetizing field $H_d$ of 1.38 T.